# Cosmic Inflation and the Arrow of Time[1]

Andreas Albrecht, UC Davis Department of Physics,
One Shields Avenue, Davis, CA 95616

**Abstract**

Cosmic inflation claims to make the initial conditions of the standard big bang "generic". But Boltzmann taught us that the thermodynamic arrow of time arises from very non-generic ("low entropy") initial conditions. I discuss how to reconcile these perspectives. The resulting insights give an interesting way to understand inflation and also compare inflation with other ideas that claim to offer alternative theories of initial conditions.

---

# 1   Introduction

One of the most obvious and compelling aspects of the physical world is that it has an "arrow of time". Certain processes (such as breaking a glass or burning fuel) appear all the time in our everyday experience, but the time reverse of these processes are never seen. In the modern understanding, special non-generic initial conditions of the universe are used to explain the time-directed nature of the dynamics we see around us.

On the other hand, modern cosmologists believe it is possible to explain the initial conditions of the universe. The theory of cosmic inflation (and a number of competitors) claims to use physical processes to *set up* the initial conditions of the standard big bang. So in one case initial conditions are being used to explain dynamics, and in the other, dynamics are being used to explain initial conditions. In this article I explore the relationship between two apparently different perspectives on initial conditions and dynamics.

My goal in pursuing this question is to gain a deeper insight into what we are actually able to accomplish with theories of cosmic initial conditions. Can these two perspectives coexist, perhaps even allowing one to conclude that cosmic inflation explains the arrow of time? Or do these two different ideas about relating dynamics and initial conditions point to some deep contradiction, leading us to conclude that a fundamental explanation of both the arrow of time and the initial conditions of the universe is impossible? Thinking through these issues also leads to interesting comparisons of different theories of initial conditions (e.g. inflation vs. cyclic models).

Throughout this article, by the "arrow of time" I mean the *thermodynamic* arrow of time. As discussed in section 3.3, I regard this to be equivalent to the radiation, psychological, and quantum mechanical arrows of time. The cosmic expansion (or the "cosmological arrow of time") may or may not be correlated with the thermodynamic arrow of time, depending on the specific model of the universe in question (see for example Hawking 1994)

I also should be clear about how I use the phrase "initial conditions". The classical standard big bang cosmology has a genuine set of (singular) "initial conditions" in the sense that the model cannot be extended arbitrarily far back in time. Much of my discussion in this article involves various ways one can work in a larger context where time, or at least some physical framework, is eternal. In that context, the problem of initial conditions that concerns us here is how some region entered into a state that reflect the "initial" conditions we use for the part of the universe we observe. This state might not be *initial* at all in a global sense, but it still seems like an initial state from the point of view of our observable universe. I will often use the term "initial conditions" to refer to the state at the *end* of inflation which forms the initial conditions for the standard big bang phase that follows. I hope in what follows my meaning will be clear from the context.

I hope this article will be stimulating and perhaps even provocative for experts on inflation and alternative theories of the initial conditions of the universe. But, In the spirit of the Wheeler volume, I've also tried to make this article for the most part accessible to a wider audience of physical scientists who may be experts in other areas but who might find the subject interesting.

This article is organized as follows: Section 2 sets the stage by contrasting a cosmologist's view of initial conditions with that of "everyone else". Section 3 presents the standard modern view of the origin of the arrow of time. In section 3.2 I discuss the case where gravity is irrelevant (which covers most everyday intuition). Then (section 3.4) I discuss the case where gravity is dominant, which allows the discussion of the arrow of time to be extended to the entire cosmos. Section 4 presents the inflationary perspective on initial conditions, and contrasts them with the perspective taken when discussing the arrow of time. Sections 5 and 6 show how these two perspectives can coexist (still with some tension) in an overarching "big picture" that allows both an explanation of initial conditions and the arrow of time. Section 7 contrasts a variety of different ideas about cosmic initial conditions in light of the insights from earlier sections. Section 8 discusses additional issues, including "causal patch" physics and problems with measures. Section 9 spells out some big open questions for the future. My conclusions appear in section 10.

## 2    The everyday perspective on initial conditions

Scientists other than cosmologists almost never consider the type of question addressed by cosmic inflation. Cosmic inflation tries to *explain* the initial conditions of the standard big bang phase of the universe. Where else does one try to explain the initial conditions of anything?

The typical perspective on initial conditions is very different. Consider the process of testing a scientific theory in the laboratory. A particular experiment is performed in the laboratory, theoretical equations are solved and the results of theory and experiment are compared. To solve the equations, the theoretician has to make a choice of initial conditions. The principle guiding this choice is trivial: Reproduce the initial conditions of the corresponding laboratory setup as accurately as necessary. The theoretician might wonder why the experimenter chose a particular setup and may well have influenced this choice, but the origin of initial conditions does not usually count as a fundamental question that needs to be addressed by major scientific advances. Instead, the initial conditions play a subsidiary role. The comparison of theory and experiment is really seen as a test of the equations of motion. The initial conditions facilitate this test, but are not typically themselves the focus of any fundamental tests.

The choice of vacuum in quantum field theories is an illustration of this point. One can propose a particular field theory as a theory of nature, but the proposal is not complete until one specifies which state is the physical vacuum. That choice determines how to construct excited states which contain particles, thus allowing one to define initial states that correspond to a given experiment.

The conceptual framework from quantum field theory has been carried over, at least loosely, into quantum cosmology where declaring "the state of the universe" based on some symmetry principle or technical definition appears to give initial conditions for the universe (Hartle and Hawking, 1983; Linde 1984; Vilenkin 1984). However, as I will emphasize in section 7, such a declaration is nothing like the sort of dynamical explanation of the initial state of the universe that is attempted by inflation, and these two approaches should not be confused with one another.

Another example where the initial conditions play a subsidiary role is in constructing states of matter whose evolution exhibits a thermodynamic arrow of time. In the next section I will discuss this example in detail before turning in section 4 to the very different perspective on initial conditions taken in inflation theory.

## 3    Arrow of Time Basics

### 3.1    Overview

In this section I discuss how to construct states of matter that exhibit a thermodynamic arrow of time. The case of systems whose self-gravity can be neglected is simplest and most well understood, so we will consider that case first. All other factors being equal, the importance of gravitational forces between elements of a system is related to the overall size of the system, and the critical size is characterized by a length scale called the "Jeans length" ($l_J$). For example, for a box of gas with size $l \ll l_J$ self gravitation is unimportant and the gas pressure can easily counteract any tendency to undergo gravitational collapse. For a larger body of gas with $l \gg l_J$ (but otherwise the same temperature, density and other local properties) the self-gravity of the larger overall mass will overwhelm the pressure and allow gravitational collapse to proceed (See for example Longair 1998).

The material in all of section 3 is pretty standard, and I give only a brief review. A much more thorough treatment of all these issues can be found in a number of excellent books on the subject (Zeh, 1992; Davies,1977) which also contain references to the original literature.

### 3.2    Without Gravity ($l \ll l_J$)

There are three critical ingredients that go into the arrow of time: Special initial conditions, dynamical trends or "attractors" that are intrinsic to the equations of motion, and a choice of coarse-graining. I will illustrate how this works in the canonical example of a box of gas.

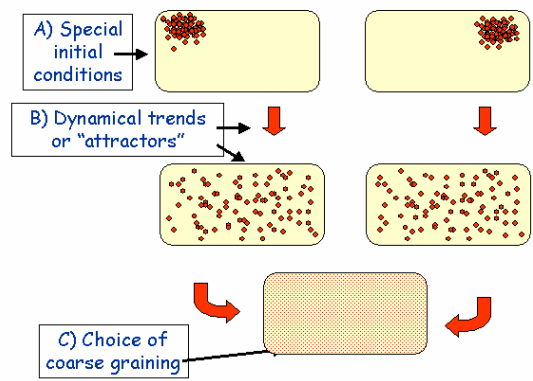

**Figure 1** A box of gas illustrates the three basic ingredients which allow the arrow of time to emerge from a fundamentally reversible microscopic world. Special out-of-equilibrium initial conditions are required, as are trends (or attractors) in the underlying dynamics drawing the system toward equilibrium. Finally, a choice of coarse-graining is essential. Without it, different initial states always evolve into different final states, and attractor-like behavior is impossible to identify.

Figure 1 illustrates a box of gas that starts in the two pictured initial states, with all the gas stuck in a corner. In each case the gas spreads out into states that look the same regardless of which corner was the starting point. This system has an arrow of time: A movie shown backwards would show a process that would never spontaneously occur in our everyday experience. Furthermore, once the gas becomes spread out, we can count on it not to spontaneously evolve back into the corner again.

Of course, according to the microscopic theory of the gas the two different initial conditions evolve into different states, and even though both look to us simply as a gas in equilibrium, the microscopic differences are retained forever, albeit in very subtle correlations among the positions and velocities of the gas particles (as well as their internal degrees of freedom). This is where coarse-graining is critical[2]. The fact that we ignore subtle differences, such as the ones differentiating the "equilibrium" states corresponding to the two different initial conditions, is the only reason we can conceive of a single stable "equilibrium state". Without course graining there would be no such thing as equilibrium, just ever-changing microscopic states. Coarse-graining is also essential to identifying the approach to equilibrium. Without coarse-graining one could only identify the microscopic evolutions of individual states, not dynamical trends, and there would be no notion of the arrow of time.

The roles of initial conditions, dynamics, and coarse-graining are closely interconnected. If the dynamics of the molecules depicted in Figure 1 were different so that, for example, the molecules were constrained to remain in the corner of the box where they started, then a typical initial state like the one depicted would *already* be in equilibrium, and such a system would not exhibit an arrow of time.

Similarly, in principle it is possible to construct formal coarse-grainings where one ignores different aspects of the microscopic state and which give arbitrarily different results. One could formally go about this, for example, by choosing some random microscopic state normally associated with equilibrium and declaring microscopic states which were dynamically nearby to that state to be in the same coarse grained "bin", and, and by similarly making other coarse grained bins from other more dynamically distant states. From that particular coarse-graining, the box of gas illustrated in Figure 1 would not exhibit an arrow of time.

---

[2] Coarse-graining is basically the act of ignoring certain aspects of microscopic states, so that many different microscopic states are identified with a single coarse-grained state (or are put in the same "coarse-grained bin"). A simple example: Take precisely defined values for position and momentum and round those values to a reduced number of digits. That gives coarse-grained coordinates in discrete phase space.

The fact that a system may or may not exhibit an arrow of time depending on the particular choice of coarse-graining creates no problems for people (like me) who are happy to see coarse-graining as a natural consequence of what kind of measurements we can actually make (something ultimately related to the nature of the fundamental Hamiltonian). However, those who wish to see the arrow of time defined in more absolute terms are concerned by the that fact that in the modern understanding the arrow of time of a given system only exists relative to a particular choice of coarse-graining and is likely to only be a temporary phenomenon (Prigogine 1962, Price 1989).

The above construction only buys you a "temporary" arrow of time because according to the microscopic theory, it *is* possible for gas in equilibrium to spontaneously evolve into one corner of its container. It just takes, on average, an incredibly long time before that happens (much longer than the age of the universe). The stability of the equilibrium coarse-grained state is deeply linked with the large number of microscopic states that are associated with the equilibrium state. From the microscopic point of view, one state is constantly evolving into another. The huge degeneracy of microscopic states associated with equilibrium means that there is lots of room to evolve from one state to another without leaving the equilibrium "coarse-grained bin".

These features are closely linked with the definition of the statistical mechanical entropy of a coarse-grained state as $\ln(N)$ (where $N$ is the number of microscopic states corresponding to the particular coarse-grained state), and with the fact that the equilibrium state is the coarse-grained state with maximum entropy. The large microscopic degeneracy of the equilibrium state is also closely related to the fact that many different initial states will all approach equilibrium.

The fact that such a large portion of all possible states for the system are associated with equilibrium, means that the special out-of-equilibrium initial states required for the arrow of time are very rare indeed. If one watched a random box of gas it would be in equilibrium almost all the time, a state with no arrow of time. At extraordinarily rare moments, there would be large fluctuations out of equilibrium and the transient associated with the return to equilibrium would exhibit an arrow of time.

In fact, due to the long periods of equilibrium the system itself has no overall time direction. The rare fluctuations out of equilibrium actually represent two back-to-back periods, each with an arrow of time pointing in the opposite direction, with each arrow originating at the point of maximum disequilibrium. Such a rare fluctuation is depicted in Figure 2.

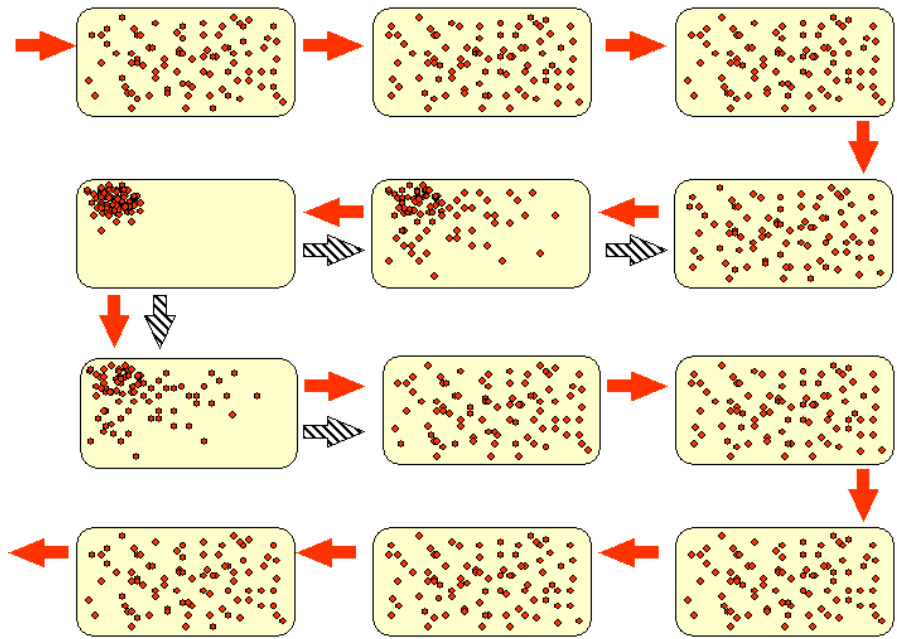

**Figure 2:** A very rare large fluctuation in a box of gas. The solid arrows depict the time series (with a randomly chosen overall direction) and the hashed arrows depict the thermodynamic arrow of time.

Interestingly, actually achieving a state of equilibrium is not absolutely necessary in order to have an arrow of time. For example, consider the generalization of the above discussion to a gas that starts out in the corner of an infinitely large box. This case is depicted in Figure 3. To achieve an arrow of time one must follow a clear dynamical trend, but a final equilibrium end point is not essential. This fact is especially relevant to the self-gravitating case discussed in Section 3.4.

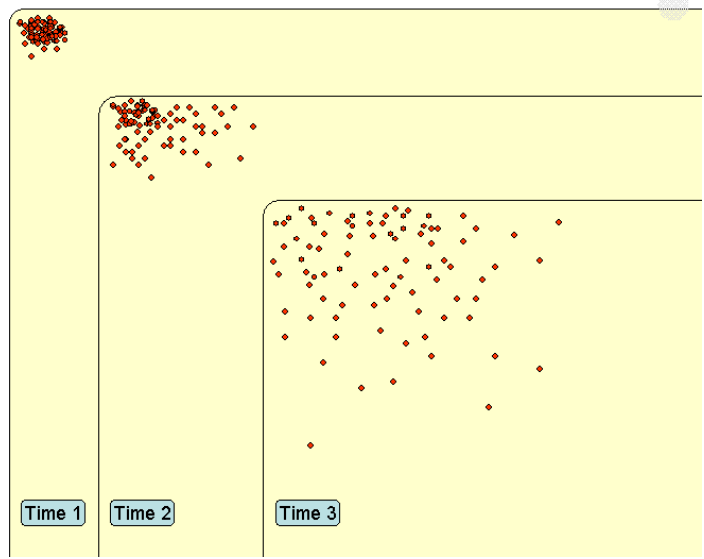

**Figure 3** A modification of the system depicted in Figure 1 to the case with an infinite box leads to a system which has a definite arrow of time, but which never achieves equilibrium.

*3.3    The key roles of the arrow of time*

The arrow of time plays a key role in many aspects of our world:

**Burning Fuel:**  The most obvious example is when we burn fuel to "produce energy" which we then harness in some way.  What we are really doing when we burn gasoline or metabolize food is producing *entropy.*  The critical resource is not the energy (which after all is conserved), but the reliability of the arrow of time.  The presence of fuel and food in our world is part of the special initial conditions that give us the arrow of time.

**Computation and Thought:** We also harness the arrow of time to make key processes irreversible. Make a mark on a page or a blackboard and you can be sure the time reverse process (the mark popping back up into the pencil or chalk) will never happen.  This allows us to make "permanent" records which are a critical part of information processing.  The use of the arrow of time for making records is ubiquitous in everyday experience, and this use of the arrow of time has also been formalized in the case of computations in work on "the thermodynamic cost of computation" (Bennett and Landauer, 1985).

Given our lack of a fundamental understanding of the process of human thought, there are many different views about the psychological arrow of time.  I personally do not expect advances in understanding human thought to bring any new insights into microscopic laws of physics (although there are probably some amazing collective phenomena to be discovered).  So I believe the psychological arrow of time is none other than the thermodynamic arrow of time, particularly as it is expressed in the making of records (or memories).

**Radiation:** A TV station can broadcast the Monday evening news with complete confidence that the radiation will be thoroughly absorbed by whatever it strikes and will not be still around to interfere with the Tuesday evening news the following day.  Furthermore, broadcasters can be confident that various absorbers will not cause interference by spontaneously emitting an alternative Tuesday evening newscast (not to mention emitting the Tuesday evening news a day early!).   The complete absence of the time-reverse of radiation absorption is understood to be one feature of the thermodynamic arrow of time in our world.  A hillside absorbing an evening news broadcast is entering a higher entropy state, and the entropy would have to decrease for any of the troublesome time reversed cases to take place.

Of course much of a radio signal propagates off into empty space.  In that case, the emptiness of the space appears to play a similar role to the infinite box in Figure 3.  Time reversed solutions, with the evening news broadcasts propagating from outer space back into the "transmitting" antenna *are* legitimate solutions to the equations of motion.  But a complete solution could not have such radiation really propagating in from infinitey.  Instead, it would have to be emitted from some astrophysical object or barring that, from the "surface of last scattering" (the most recent point in the history of the universe when the universe was sufficiently dense to be opaque).  Any of these astrophysical or cosmological sources would have to be in a much lower entropy state than we expect if they are to produce time reversed "evening news" radiation.  So in the end, the radiation arrow of time is none other than the thermodynamic arrow of time, which is the topic of this article. I should note that much of our understanding of the radiation arrow of time was developed by John Wheeler (Wheeler and Feynmann, 1945, 1949) who we honor with this volume.

**Quantum Measurement:** An arrow of time is critical to quantum mechanics as we experience it. Once a quantum measurement is made there is no undoing it, and one says the wavefunction has "collapsed". There are different attitudes about this collapse. One approach is to see this collapse as a consequence of establishing stable correlations: A double slit electron striking a photographic plate is only a good quantum measurement to the extent that the photographic plate is well constructed, and has a very low probability of re-emitting the electron in the coherent "double slit" state. Good photographic plates are possible because of the thermodynamic arrow of time: The electron striking the plate puts the internal degrees of freedom of the plate into a higher entropy state, which is essentially impossible to reverse. Furthermore, different electron positions on the plate become entangled with different states of the internal degrees of freedom, so there is essentially no interference between positions of the electron. From this point of view (which I prefer) the quantum mechanical arrow of time is none other than the thermodynamic arrow of time[3]. Others want to establish a quantum arrow of time that is separate from the thermodynamic arrow, but no well-established theory of this type exists so far.

### 3.4    With Gravity ($l \gg l_J$)

When the self-gravity of a system is significant, the dynamical trends are very different. While the gas in the box discussed in Section 3.2 tended to spread out into a uniform equilibrium state, for gravitating systems the trend is toward gravitational collapse into a state with less homogeneity. Interestingly, when gravitational collapse runs its course, matter also approaches a kind of equilibrium state: the black hole. As is fitting for equilibrium states, one can even define the entropy of a black hole, namely the Beckenstein-Hawking entropy given by

$$S_{bh} = 4\pi M^2 \qquad (1)$$

for a black hole of mass $M$. Although black hole entropy is not as well understood as the entropy of a box of gas, it certainly fits with the general picture quite well[4].

As with any other system, gravitating systems will exhibit a thermodynamic arrow of time if they have special "low entropy" initial conditions. The observed universe is an excellent example. The observed universe certainly has a sufficiently strong self-gravity so as to be subject to gravitational collapse (namely $l \gg l_J$). But this trend is in its very early stages, and is very far from having run its course. That is, the observed universe is very far from forming one giant black hole. Penrose (1979) quantified this fact by comparing the entropy of the very early universe (as measured by the ordinary entropy of the cosmic radiation fluid) with the entropy of a black hole with mass equal to the mass of the observed universe. The result is that the entropy of the early universe is 35 orders of magnitude smaller than the maximal entropy black hole state:

$$S_{Univ} \approx 10^{-35} S_{bh-Max} = 10^{-35} 4\pi M_{Univ}^2 . \qquad (2)$$

---

[3] Significant contributions to this perspective come from John Wheeler and his students (Everett, 1957; Wheeler and Zurek 1983; Zurek 1991). See also Albrecht (1992, 1994).

[4] This discussion is classical and does not include the effects of Hawking radiation. Including Hawking radiation might make it more difficult to formulate an "ultimate equilibrium" state for general gravitating systems, but that does not matter for the discussion here. Hawking radiation is irrelevant on the temporal and spatial scales over which gravitational collapse defines the arrow of time in the observed universe.

As Penrose originally argued, the low entropy of the early universe is the ultimate origin of the arrow of time we experience. Just as the box of gas depicted in Figure 1 evolves reliably toward a more homogeneous state, giving the system an arrow of time, so too does the universe follow its own dynamical trends from a state of homogeneity toward a state of gravitational collapse. In the case of the universe, it is not clear that a final equilibrium black hole state will be achieved, so it may turn out that the better analogy is the infinite box of gas depicted in Figure 3. The key point is that the universe starts out in a very special state which is far from where the dynamical evolution wants to take it. The realization of this evolution results in an arrow of time.

I conclude this section with an illustration of the relationship between the arrow of time of the universe as a whole, as expressed by a trend from homogeneity toward gravitational collapse, and the simple everyday examples of the arrow of time as discussed in Section 3.2. Figure 4 illustrates a process by which we might construct a box of gas with an arrow of time such as that depicted in Figure 1. The gas is pumped into the corner by an electric pump, with electricity generated by fossil fuels. The organic matter which formed the crude oil that was refined into the fuel was created by photosynthesis which harnessed the sun's radiation. The hot sun radiating into cold space is our local manifestation of the ongoing process of gravitational collapse throughout the universe.

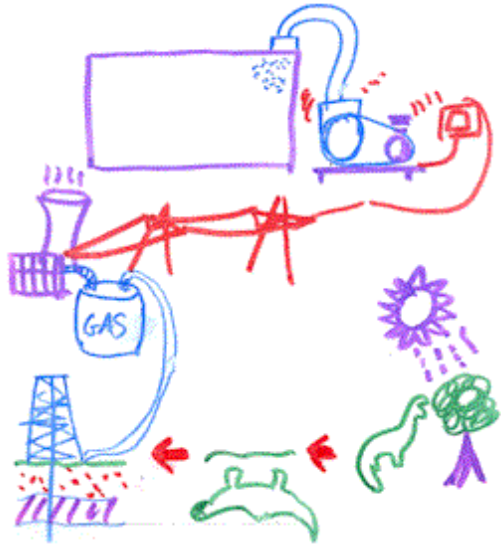

**Figure 4** One can pump on a box of gas to move all the gas into one corner. When the pumping stops the gas spreads out, exhibiting an arrow of time as depicted in Figure 1. In this example the pump uses electricity generated by fossil fuels, produced from organic matter which originally harnessed solar energy to be created. The hot sun radiating into cold space is our local manifestation of the ongoing process of gravitational collapse throughout the universe. This example illustrates the links between everyday examples of the arrow of time and the overall arrow of time of the universe, as expressed through gravitational collapse.

As discussed in Section 3.3, what we traditionally call sources of power or energy are really sources of *entropy*, which allow us to harness the arrow of time.  Most of our power sources can be traced to radiation from the sun, as in Figure 4.  The exceptions are geothermal power (which harnesses the gravitational collapse that produced the earth itself) and nuclear fission power (which uses unstable elements produced in the collapse of stars other than the sun).  Fusion energy exploits another sense in which the universe is out of equilibrium: The homogeneous cosmic expansion proceeds too quickly for the nuclei to equilibrate into the most stable element, and instead produces nuclei which are out of chemical equilibrium (i.e. not in the most tightly bound nuclei).  This leaves an opportunity to release entropy by igniting fusion processes that bring nuclear matter closer to chemical equilibrium[5].  This issue (and its links to the initial state of the universe) will be discussed further in section 4.2.

## 4    Cosmic Inflation: preliminaries

### 4.1    The inflationary perspective on initial conditions

In the previous section we discussed how the thermodynamic arrow of time must necessarily be traced to special initial conditions.  In particular, we discussed how the overall arrow of time in the universe is linked to special initial conditions for the universe that are far away from the dynamical trend toward gravitational collapse. With this understanding, one can accept these special initial conditions in the usual subsidiary role: Experimental data tells us that the universe has an arrow of time, so to model the universe we obviously must choose initial conditions appropriately.  To this end, the homogeneous and isotropic expanding initial conditions of the standard big bang are a great choice, and they do indeed (when combined with a suitable initial spectrum of small primordial perturbations) give an excellent match to all the observations.

But enthusiasts of cosmic inflation (Guth, 1981; Linde 1982; Albrecht and Steinhardt 1982) take a very different view.  Typical presentations of cosmic inflation start by presenting a series of cosmological "problems" that appear to be present in the standard big bang (see for example (Guth, 1981) or (Albrecht 1999)).  Many cosmologists were concerned about these problems even before the discovery of inflation.  The first two of these problems (the "Flatness" and "Homogeneity" problems) basically state that the initial conditions are far removed from the direction indicated by the dynamical trends.

The flatness problem is the observation that the dynamical trend of the universe is away from spatial flatness, yet to match today's observations, the universe must have been spatially flat to extraordinarily high precision.

The homogeneity problem is exactly a re-statement of the main point of Section 3.4 of this article:  The universe is in a state far removed from where gravitational collapse would like to take it.  The discussion in Section 3 emphasized that a property of this sort is absolutely required in order to achieve an arrow of time.  From that point of view, the special initial conditions of the universe are not a puzzle, but the answer to the question "where did the arrow of time come from?".

---

[5] The sun and other stars in fact produce an interesting combination of "gravitational collapse power" enhanced by nuclear fusion power.

However, most cosmologists would instinctively take a different perspective. They would try and look further into the past and ask how could such strange "initial" conditions possibly have been set up by whatever dynamical process went before. Since the initial conditions are counter to the dynamical trends, it seems on the face of it that the creation of these initial conditions by dynamics is a fundamental impossibility. In fact, the "horizon problem" adds to this dilemma by observing that there was insufficient time in the early universe for causal processes to determine the initial conditions of what we see, even if somehow there was a way to fight the dynamical trends.

So we have two different points of view. An inflationist wants the initial conditions of the universe to be more natural, but the intellectual descendants of Boltzmann would say they had better *not* appear natural: The unnaturalness of initial conditions is precisely what is necessary in order to have an arrow of time, so it appears the price of "natural" initial conditions is the absence of an arrow of time. In addition, considering the general comments from section 2, the inflationist would seem to be in a weaker position. Certainly the strongest tradition in physics is for initial conditions to play a subsidiary role, unquestioningly assigned whatever form is required for the situation at hand.

The goal of this article is to reconcile these points of view. To get started, I will give two illustrations of familiar situations where the initial conditions do play a more critical role, and for which dynamics actually creates special initial conditions. With the lessons learned from these illustrations, we will be ready to scrutinize cosmic inflation.

### 4.2    Illustration 1: Big Bang Nucleosynthesis

One of the classic results from cosmology is the synthesis of nuclei in the early universe. Using cross-sections determined in laboratories on earth, one can calculate the cosmic abundances of different nuclei at different times. As the universe cools and becomes more dilute, a point is reached were the mass fraction stops changing. These "frozen out" values are the predictions of "primordial nucleosynthesis".

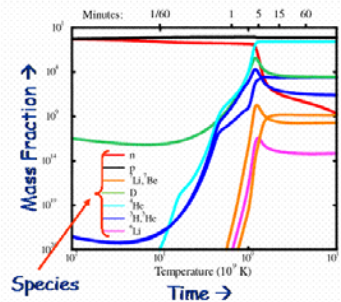

**Figure 5** The evolution of nuclear species in the early universe: The mass fractions freeze out at specific values, leading to predictions of nuclear abundances from early universe cosmology. Figure adapted from Burles et al. (2001).

One might very well wonder whether it is really possible to make such predictions. Surely the state at late times depends on what you chose for initial conditions. Would it not be possible to get any prediction you want by choosing suitable initial conditions? In fact, it turns out that the predictions *are* almost entirely independent of the choice of initial conditions: Any initial conditions for the nuclear abundances are erased by the drive toward nuclear statistical equilibrium, which *sets up* the "initial conditions" for the subsequent evolution. Figure 6 illustrates this effect with different initial conditions.

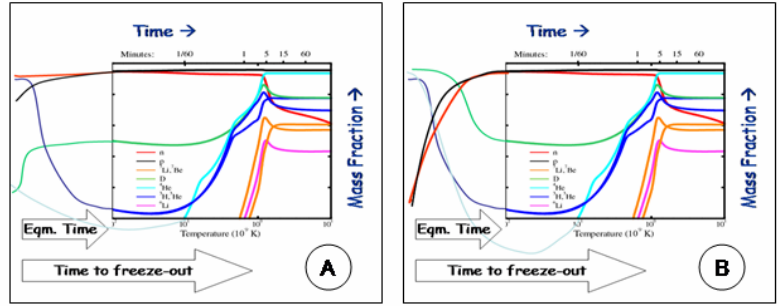

**Figure 6** Different initial mass fractions of nuclei would all rapidly approach nuclear statistical equilibrium, resulting in the same "initial" conditions for the subsequent process of nucleosynthesis. Thus for nucleosynthesis, the initial conditions are determined by the dynamics of a previous epoch. Figure adapted from Burles et al. (2001). (Curves outside the box of the original plot are sketched in, and do not represent actual calculations.)

As discussed in section 3, Different initial conditions always do evolve into different states, when viewed at the microscopic level. The two cases in Figure 6 approach the same equilibrium state only because one is coarse-graining out subtle correlations among particles that carry information about the initial conditions. This coarse-graining is implemented by focusing just on the mass fractions, which represent just a small amount of information about a microscopic state.

So big bang nucleosynthesis is an example of a situation where "initial" conditions definitely do not play a subsidiary role, but are critical to any claim that one is actually making predictions. In this case, the dynamics of an earlier epoch (namely the approach to equilibrium) step in to set up the subsequent initial conditions, just as one hopes cosmic inflation can set up initial conditions for the big bang cosmology.

### 4.3   Illustration 2: Gas in a block of ice

I now discuss an even simpler example, which will turn out to be conceptually very similar to the nucleosynthesis case. Consider two boxes of gas similar to those depicted in Figure 1, but now supposed they are encased in blocks of ice, as shown in Figure 7.

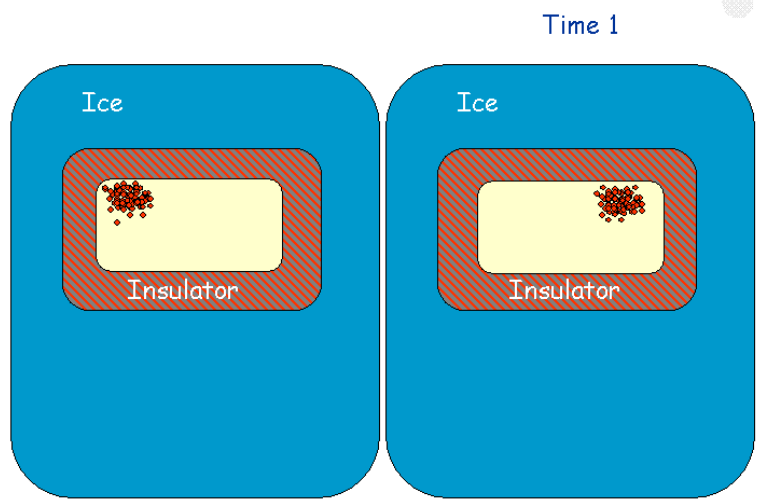

**Figure 7** Boxes of gas encased in ice

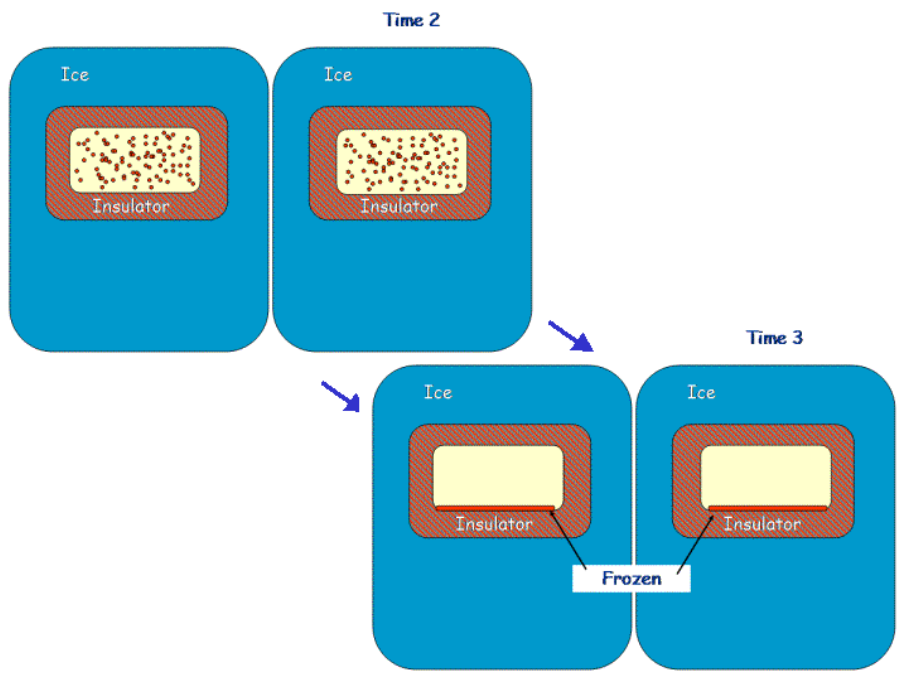

**Figure 8** Subsequent evolution of the system depicted at "Time 1" in Figure 7. The gas inside the box equilibrates first, setting up the initial conditions for the subsequent condensation and freezing.

The insulator between the ice and the gas is not perfect, but serves to slow down the equilibration time between the two. Figure 8 illustrates the subsequent evolution. The gas has plenty of time to equilibrate, and the equilibration sets up "initial" conditions for the process of condensing and freezing. As a result, the uniform state of the frozen gas at "Time 3" can be *predicted*.

## 4.4    Equilibrium and de Sitter space

Both of the above illustrations used an early period of equilibration to set up initial conditions for subsequent evolution. To understand how this concept carries over to the case of inflation, we first have to expand on the discussion of section 3.4, where equilibrium was discussed for gravitating systems.

Einstein first proposed the "cosmological constant" (known as Λ) early in the days of general relativity. Later, it was realized that certain scalar fields can at least temporarily enter a "potential dominated state" which closely mimics the behavior of a cosmological constant. A cosmological constant, roughly speaking, acts like a *repulsive* gravitational force, and Einstein first proposed it to balance the normal attractive force of gravity in order to model a static universe. However, that idea did not work because such a balance is not stable. The natural evolution of matter in the presence of a non-zero cosmological constant eventually becomes dominated by the repulsive force and is driven apart exponentially fast by the resulting expansion. The expansion rate is

$$H = \sqrt{\frac{8\pi G}{3}\Lambda} \; .$$

(3)

After waiting out a suitable "equilibration time" the exponential expansion will empty out any given region of the universe, leaving nothing but the cosmological constant (or potential dominated matter) which does not dilute with the expansion. This exponentially expanding empty (but for the cosmological constant) spacetime is known as de Sitter space. The approach to de Sitter space is essentially the opposite of the gravitational collapse discussed in section 3.4: Instead of approaching an equilibrium state of total gravitational collapse (a black hole), with a non-zero cosmological constant the universe asymptotically approaches de Sitter space, a state of essentially total "un-collapse".

As we discussed in section 3, it seems natural to associate the notion of equilibrium with endpoint states toward which many states are dynamically attracted. This perspective makes it natural to think of a black hole as an equilibrium state, and thus it seems natural to define black hole entropy. Perhaps not surprisingly, similar arguments to the black hole case produce a definition of the entropy of de Sitter space (Gibbons and Hawking 1977):

$$S_{dS} = \frac{3\pi}{\Lambda} \; .$$

(4)

The statistical foundations of de Sitter space entropy are probably even more poorly understood than the black hole entropy, but it certainly fits in nicely with the heuristic notion of entropy and equilibrium considered here. Also, the part of de Sitter space toward which a cosmological constant dominated universe evolves is homogeneous and flat: two features of the big bang cosmology that inflation seeks to explain.

## 4.5   The potential dominated state

Models of cosmic inflation use a scalar field $\varphi(x)$ (the inflaton) to mimic the behavior of a cosmological constant for a certain period of time. This cosmological constant-like behavior is achieved when the inflaton is in a potential dominated state. Specifically, it is the inflaton stress energy, given by an expression like

$$T_{\mu\nu} = \partial_\mu \partial_\nu \varphi(x) + g_{\mu\nu}\left[ g_{\alpha\beta}\partial_\alpha \partial_\beta \varphi(x) + V(\varphi)\right] \xrightarrow{g\partial\partial\varphi \ll V} g_{\mu\nu}V(\varphi) \qquad (5)$$

that must be dominated by the potential term $\propto V(\varphi)$ for the inflaton to look like a cosmological constant[6]. So ultimately constraints like

$$V(\varphi) \gg g_{\alpha\beta}\partial_\alpha \partial_\beta \varphi(x) \qquad (6)$$

must hold. As has been known since the early days of inflation, and has been emphasized over the years (Penrose 1989, Unruh 1997, Trodden and Vachaspati 1999, Hollands and Wald 2002), the potential dominated state for the inflaton is a very special state. The field $\varphi(x)$ has a huge number of degrees of freedom, and many possible states of excitation. Only a tiny fraction of these will obey the constraints in Eqn. (6) sufficiently strongly to allow the onset of inflation. This fact will be important in what follows.

# 5   Cosmic inflation

## 5.1   Basic inflation

The basic idea of inflation is that the universe entered a potential dominated state at early times. If the potential dominated phase was sufficiently long, they spacetime would have had a chance to equilibrate toward de Sitter space[7]. The de Sitter space has the flatness and homogeneity properties required for the early stages of the big bang, so via the approach to de Sitter space these features are acquired dynamically[8].

But of course in the early stages of the big bang the universe is full of ordinary matter, not potential dominated matter. A critical part of cosmic inflation is *reheating*: After a sufficient period of inflation the potential dominated state decays (or "reheats") into ordinary matter in a hot thermal state.

In typical modern inflation models the instability is a classical one of the "slow roll" type illustrated in Figure 9. The critical degree of freedom driving inflation is the homogeneous piece (or average value) of the inflaton field $\varphi$, depicted in the figure. This degree of freedom can be thought of as "rolling" in its potential $V(\varphi)$. At the onset of inflation $\varphi$ starts out in a relatively flat part of $V(\varphi)$ so the small values of the time derivative $\partial_0 \varphi$ (required for Eqn. (6) to hold) can be maintained. The field is rolling slowly here, and the potential domination causes exponential expansion to set in.

---

[6] $\partial_\mu \varphi(x)$ denotes the space and time derivatives of $\varphi(x)$. For further background see for example Kolb and Turner (1999)

[7] The fact that inflation is never in perfect equilibrium has been analyzed by Albrecht *et al* (2002)

[8] The standard big bang models the universe all the way back to an initial singularity of infinite density and temperature. Inflation provides an alternate account of the very early universe, and matches on to the standard big bang at a finite time after the initial big bang singularity. The question of whether other singularities necessarily precede inflation is under active investigation, see "Eternal Inflation" in section 7.

However, $\varphi$ is never completely stationary, and it eventually reaches a part of the potential that is steeper. At that point $\varphi$ speeds up, Eqn. (6) no longer holds, and the exponential expansion is over. If $\varphi$ is suitably coupled to ordinary matter, energy can couple out of $\varphi$ and into ordinary matter in the non-slow-roll regime, creating the right conditions for the beginning of the standard big bang.

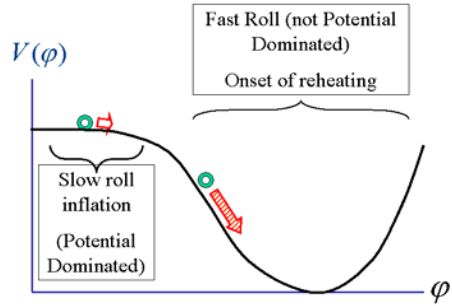

**Figure 9** The homogeneous piece of the inflaton field (depicted here) controls inflation by first rolling slowly in a flat part of its potential $V(\varphi)$ (allowing potential domination and exponential expansion) and then entering a steeper part of the potential that ends the slow roll and allows reheating to occur.

Quantum corrections to the above discussion allow one to predict deviations (or perturbations) from perfect homogeneity produced during inflation, which evolve into galaxies and other structure in the universe. These are discussed further in section 6.3.

At this stage we do not have a strongly favored "standard model" for the inflaton. There are a huge number of workable proposals for the origin of the scalar field, $V(\varphi)$, etc., but no clear favorite and none that are deeply rooted in well-established theories of fundamental physics. This fact must be regarded as a weakness in the inflationary picture. But to be fair, that situation might be more a reflection of our generally primitive understanding of fundamental physics at the relevant high energy scales rather than anything intrinsically suspect about inflation.

### 5.2    Inflation and the arrow of time

We now have seen three examples where a process of equilibration generates dynamically predicted initial conditions for the next stage of evolution. (The examples are Big Bang Nucleosynthesis, the gas in ice, and cosmic inflation.) But how do these examples address the key question of this article, namely how can one harness equilibration to make a special initial condition "generic" and still have an arrow of time (that is, non-generic initial conditions)?

In each of these examples the question is resolved in the same way. The equilibration during the first "initial condition creating" stage is not equilibration of the entire system, but just of a *subsystem*. In each case there are additional degrees of freedom which are never in equilibrium that drive the system and carry information about the arrow of time.

In the case of Big Bang Nucleosynthesis, it is the spacetime (or gravitational) degrees of freedom that are out of equilibrium. The universe is not one giant black hole (equilibrium for a normal gravitating system) but rather a homogeneous and isotropic expanding Big Bang state (which, as Penrose taught us, is far out of equilibrium). Against this background, the nuclear reactions are able to maintain chemical equilibrium among nuclear species at early times, when the densities and temperatures are high. As the out-of-equilibrium degrees of freedom (namely the cosmic expansion) cool the universe, the lower energies and densities put matter in a state that can no longer maintain nuclear statistical equilibrium. The expanding spacetime background is the out-of-equilibrium subsystem that drives the change, first allowing the "nuclear species subsystem" to enter chemical equilibrium and then (having thus set up the "initial conditions") driving the nuclear matter toward the out-of-equilibrium conditions that produce the predicted mass fractions from primordial nucleosynthesis.

For the ice and gas (Figure 8), the ice and gas start far from equilibrium but with a slow equilibration time due to the presence of the insulator. The initial equilibration (of gas within the box) is just the equilibration of the gas subsystem. Viewed as a whole, the ice and gas are still out of equilibrium, even as the gas subsystem spreads out into an equilibrium state within its box (which of course defines the "initial" conditions for what comes next).

For inflation, the *inflaton field* is the out of equilibrium degree of freedom that drives other subsystems. The inflaton starts in a fairly homogeneous potential dominated state which is certainly not a high entropy state for that field (Trodden and Vachaspati 1999). In a well-designed inflation model the special potential dominated inflaton state "turns on" an effective cosmological constant and leaves it on for an extended time period, allowing plenty of time for the matter to equilibrate toward de Sitter space. But the slow roll inflaton instability eventually "turns off" the cosmological constant, and the continued out-of-equilibrium evolution of the inflaton leads to a period of re-heating followed by conditions appropriate for the early stages of the standard big bang (at which point the spacetime is the out-of-equilibrium degree of freedom driving the subsequent arrow of time).

So while inflation does dynamically "predict" the special initial state of the big bang phase, it does not *predict* the arrow of time. Inflation "passes the arrow of time buck" to the special initial conditions of the inflaton field. An arrow of time, by its fundamental nature, requires non-generic initial conditions. For a big bang universe created by inflation, the non-generic quality of the initial conditions that gives us an arrow of time can be traced right back to the special inflaton initial state.

To better understand the role of inflation and its relationship to the arrow of time, it is necessary to put the above discussion in a larger context. That exercise (which is the subject of the following section) will help us understand how inflation has a crucial role, despite the fact that that it does not *predict* every aspect of the universe we observe. (N.B. you can find an early discussion of some of the key issues from this section and the next in a series of papers by Davies (1983,1984) and Page 1983.)

## 6    Initial Conditions: The Big Picture

### 6.1    *Data, theory, and the "A" word*

To what extent should we use observational data to confront theoretical predictions, and to what extent should that data instead be used as input to theoretical models, in order to constrain free parameters? The debate about this issue can get extremely passionate, and often involves using "the A word" or the "anthropic principle".

I believe that the reality behind the passions is pretty straightforward, and offers clear guidance about how to proceed: Every theory known so far requires some observational data to be used as input, to constrain charges, masses, and other parameters. On the other hand, pretty much everyone would agree that if a new theory required less data as input (i.e. had fewer free parameters to set) and could in turn predict some of the data that the old theory used as input, then the new theory would simply be better than the old theory, and would supercede it.

A consequence of this line of reasoning is that data should be treated as a precious resource: Using data up to set parameters should be avoided if at all possible. It is much better to save up the data to use to test the predictions of your theory after a minimal number of parameters are set. If you are sloppy about this issue, your most serious penalty is not really the harsh criticism you might experience at the hands of physicists with other passionately held views. The real threat is that another approach that is more efficient with the data could simply leave your line of thinking behind in the dust.

So while "pro anthropic" scientists tend to alarm their colleagues by apparently freely using up precious data to constrain models[9] the anthropicists might be equally indignant that many of their opponents seem unwilling to acknowledge that some data really does need to be used up as input.

A more fruitful approach lies between these two extremes: Admit that some of our precious data needs to be used up as input, but work as hard as possible to use up as little data as possible in this manner.

*6.2    Using the arrow of time as an input*

The arrow of time, as it is currently understood, simply has to be used as an "input" to any theory of the universe. At its most fundamental level, the arrow of time emerges from evolution from a special initial state toward more generic subsequent states (where "generic" and "non-generic" are defined relative to the natural evolution under the equations of motion and also relative to a particular coarse-graining). To have an arrow of time, there must be something non-generic about the initial state. That property of the initial state must be chosen, not because it is a typical property but because that (necessarily atypical) property is required in order to have an arrow of time.

An attractive way of incorporating this line of reasoning into a "big picture" follows up on the discussion of Figure 2 in section 3.2. Figure 2 shows a random large fluctuation in an "equilibrium" box of gas creating conditions where there temporarily is an arrow of time. If one thinks of a box of gas sitting there for all eternity, such events, although rare, will occur infinitely many times. In this kind of picture, the special initial conditions that produce an arrow of time are not imposed on the whole system at some arbitrary absolute origin of time. Instead the special "initial" conditions are found by simply waiting patiently until they occur randomly. Boltzmann (1897, 1910) already was thinking along these lines a hundred years ago, but found some aspects of this argument deeply troubling.  In section 6.4 we will discuss Boltzmann's problem, and see what inflation has to say about it.

---

[9] Statements that life could not exist without some detailed property of the known physical world come across as gratuitous to many physicists (including me), since we really do not have a clue what great varieties of "life" might be possible. Without some more concrete expression of this idea, one appears to be simply using the physical property as input, and giving up on actually predicting it.

Most modern thinking about inflation borrows at least some aspects from Figure 2. One typically imagines some sort of chaotic primordial state, where the inflaton field is more or less randomly tossed about, until by sheer chance it winds up in a very rare fluctuation that produces a potential dominated state (Linde, 1983).  One important difference between the box of gas and the "pre-inflation" state is that it is much easer to calculate things for the box of gas.  Although very interesting pioneering work has been done (see for example Linde 1996), we still do not appear very close to a concrete systematic treatment of a chaotic pre-inflation state.

Of course once it is possible to create a period of inflation, one may not need to know too much about the pre-inflation state.  In many models, inflation creates such a large volume of the universe in the inflated domain that the predictions appear to be insensitive to many details of the pre-inflation state.  Still, one certainly needs to know enough about the pre-inflation state to at least roughly establish such insensitivity.

But there is an even bigger question lurking behind this issue.  If one is willing to concede that even with inflation, special initial conditions must ether be stumbled upon accidentally or imposed arbitrarily, what role is left for inflation?  Why not simply wait around for the big bang itself to emerge directly out of chaos, or impose big bang initial conditions directly on the universe, without bothering with an initial period of inflation[10]? (See for example Barrow 1995.) The answer is that even though inflation is not all-powerful, and cannot create the big bang from absolutely anything, inflation still has a great deal of predictive power which allows one to make more economical use of the data than one could in the absence of inflation.

### 6.3    Predictions from cosmic inflation

Once the special inflaton initial conditions get inflation started, a whole package of predictions are made.  The universe is predicted to be homogeneous, with a density equal to the critical density (to better than 0.01% accuracy).  A spectrum of perturbations away from perfect homogeneity are also predicted, with a specific "nearly scale invariant" form.  Perhaps most important for this discussion, the volume of a typical region that has these properties is huge, exponentially larger than the entire observed universe.  These predictions go well beyond the basic notion of what the standard big-bang cosmology describes.  Taking the standard big bang model on its own, there is no particular reason to expect the density to be nearly critical, or to expect a particular form for the spectrum of perturbations.  Currently, a large body of data supports the inflationary predictions (see for example Figure 10 as well as Albrecht (2000) for examples and more information).

---

[10] This question is raised directly by Barrow (1996) and is also closely related to other concerns (Penrose 1985, Unruh 1997, Linde et al. 1994, Linde et al. 1996, Vanchurin et al. 2000, Wald 2002).

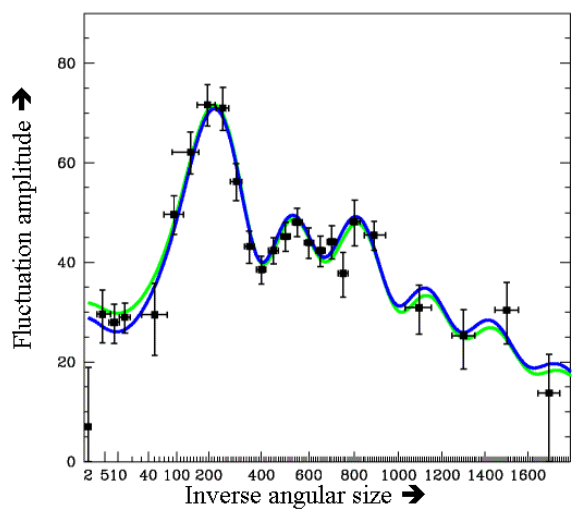

**Figure 10** This figure from (Tegmark and Zaldarriaga 2002) shows a compilation of measurements of anisotropies in the Cosmic Microwave Background (points) along with curves from inflation models as a function of inverse angular scale on the sky. The left-right location of the peak structure is very sensitive to the overall density of the universe. Current best estimates show the density is consistent with the critical value predicted by inflation to within error bars around 10% (Wang et al. 2001). The oscillatory behavior and the lack of an overall sharp rise or drop across the plot also support the predictions of inflation (Wang et al. 2001, Albrecht 2000).

No inflation model predicts that the entire universe is converted to a big-bang like state. In many models, quantum fluctuations take the inflaton back "up the hill" sufficiently frequently that at any time after inflation starts, regions that are still inflating actually dominate the volume of the universe (Linde 1986)). But if you do find a region with ordinary matter (as opposed to the potential dominated inflating state) that region will be exponentially large, and have the properties described in the previous paragraph[11].

Inflation is best thought of as the "dominant channel" from random chaos into a big bang-like state. The exponentially large volume of the big bang-like regions produced via inflation appear to completely swamp any other regions that might have fluctuated into a big-bang like state via some other route. So if you went looking around in the universe for a region like the one we see, it would be exponentially more likely to have arrived at that state via inflation, than some other way, and is thus strongly predicted to have the whole package of inflationary predictions.

### 6.4 Boltzmann's "efficient fluctuation" Problem

It is pretty exciting to have a theory of initial conditions with plenty of specific predictions to test. The fact that inflation offers such a theory has had a huge impact on the field of cosmology, and has motivated high ambitions on both the theoretical and observational sides of the field. But inflation also addresses another issue that had Boltzmann worried a century ago.

---

[11] Note however that here I am using up additional data as input.

Boltzmann also was trying to think of the "dominant channel" into a universe like ours, but without the benefit of inflationary cosmology. Boltzmann realized that the only way one could expect nature to produce the unusual "initial" conditions that lead to an arrow of time is to wait for a rare fluctuation of the sort depicted in Figure 2. But that "dominant channel" also comes with its package of somewhat disturbing predictions. In particular, rare fluctuations in ordinary matter seem to be very stingy about producing regions with an arrow of time. If you use as input the data that you are sitting in a room, like whatever room you are sitting in, and that it has existed for at least an hour, then by far the most likely fluctuation to fit that data is a room that fluctuates alone in the midst of chaos, and is immediately destroyed by the surrounding chaos as soon as the hour is up. If you want to look for a larger piece of your world (the whole building which contains your room, the whole city, the whole planet, etc.) you would have to wait around for an even more rare fluctuation, and by far the most likely fluctuation would just barely fit the input data, and exhibit utter chaos everywhere else.

So Boltzmann (and many others since) worried that if our world really emerged from a random fluctuation, then a strong prediction is made that we exist in the midst of utter chaos. The fact that instead we live in a universe billions of lightyears in size which is extremely quiet and un-chaotic, and that seems to have room for not just our cozy planet, but many more like it seems to be in blatant contradiction to these predictions. To get a rare fluctuation to produce all that, you would have to use up all those features of the universe as input data. None of those features would be predicted. (For a nice account of this issue, see section 3.8 of Barrow and Tipler (1986).)

Cosmic inflation gives a very attractive resolution to this problem. The big picture is similar, in that one has to wait for a rare fluctuation to create the universe we observe. But inflation says the most likely rare fluctuation to produce the world we see is not the random assembly of atoms, molecules and larger structure directly out of a chaos of ordinary matter. Inflation offers a completely different set of dynamics, where a small fluctuation in the inflaton field gives rise to regions that look like our universe, but which actually generically extend exponentially further beyond what we see. Inflation transforms the large scale nature of our universe from a mystery into a prediction.

## 7   Comparing different theories of initial conditions

My discussion has emphasized four key aspects of the inflationary picture:

1) <u>Attractor:</u> Inflation exhibits "attractor" behavior (or equilibration toward de Sitter space) which causes many different states to evolve into states that resemble the early stages of the big bang.

2) <u>Volume factors:</u> Inflation generates exponentially large volumes, which gives extra weight to the inflationary channel into these early big bang states.

3) <u>Arrow of time:</u> Despite features 1) and 2) the initial conditions for inflation need to be non-generic to some degree. This is required in order to have an arrow of time.

4) <u>Predictions:</u> Still, when points 1-3 are taken together, inflation produces an impressive package of predictions which overall allow one to use up less data as input than one would have to do in the standard big bang model taken alone.

Today there are a variety of different ideas about initial conditions in play, and it is interesting to consider how different ideas compare with inflation on these four key points:

**Chaotic Inflation:** The discussion in this article embraces the ideas put forth by Linde on chaotic inflation[12]. This article should be seen as a further extension of these ideas.

**Eternal Inflation**: There has been a lot of discussion recently about whether it is possible to describe the universe as an eternal inflating state with (exponentially large) islands of reheated matter. This description would allow one to forget about trying to understand the "pre-inflation" state altogether There simply would be no pre-inflation. Different viewpoints have emerged on this subject. One view states that such an eternally inflating state is impossible to create because singularities necessarily arise. These singularities can take a variety of forms, but in each case the upshot is that additional initial data is required, implying some notion of "pre-inflation". (Borde et al. 2001)

Another view is that the very statement that one is looking for an eternally inflating state contains enough information to resolve such singularities. Aguirre and Gratton (2002) claim that when one uses this information to good effect, there is one obvious choice for the "pre-inflation" state. If that choice is made, Aguirre and Gratton argue that a global state is constructed which can, in the end, be thought of as defining an eternally inflating state. The eternally inflating state that emerges from that approach has specific global properties that reflect an arrow of time. In particular, an array of regions of reheating (or decay of the potential dominated state) must be organized coherently to be pointing in a commonly agreed "forward direction". In fact, there are actually two different "back to back" coherent domains in this picture, with arrows of time pointing in opposite directions. The coherence must extend over infinitely many reheated regions, distributed throughout an infinitely large spacetime volume.

Several technical issues remain unanswered (for example whether the construction of Aguirre and Gratton can be implemented at the level of full fundamental equations), but here I simply comment on how these two perspectives relate to the four key points mentioned above. I start with the Aguirre-Gratton picture: 1) The eternal inflation picture specifically avoids needing attractors. By fiat the state of the universe is specified completely, and there is no need to draw other states toward it. 2) In the Aguirre-Gratton picture there is only one way to create big bang-like regions, so although the exponentially large volume factors certainly are present, they do not seem to have as crucial a role as they have in a more standard inflationary picture. 3) In the Aguirre-Gratton construction, the arrow of time is put in by hand. One simply declares "the universe is in this state", and it happens to have an arrow of time. The only conceptual difference between the Aguirre-Gratton idea and simply declaring "the universe is in a standard big bang state" (in other words, forgetting about inflation altogether) is the claim (still debated) that the eternal model does not have singularities. The Aguirre-Gratton idea specifically tries to eliminate the role of a rare random fluctuation of the inflaton that one sees in the standard discussions of chaotic inflation (and replaces it with a special choice of state for *all* time).

On the other hand, Borde et al. (2001) say that singularities exist which make it impossible to extend the inflationary state eternally back in time. This perspective fits perfectly with the picture developed in sections 5 and 6 of this article, where the singularity is resolved by extending back in time not with more inflation, but into some more chaotic state of spacetime and matter (probably with its own naturally occurring singularities that need to be resolved by a more fundamental theory).

---

[12] For a nice overview, with references to the original literature see (Linde 1997).

**The Ekpyrotic Universe:** This idea basically suggests a way of extending the story of the universe backward past the big bang phase into an epoch where the universe can be described (presumably at a more fundamental level) by colliding "branes" in a higher dimensional space (Khoury et al. 2001a). 1) The proposed dynamics do *not* contain any attractor behavior 2) nor do they have any exponential volume creation. 3) The arrow of time and many other features of the big bang cosmology are a direct consequence of very special properties of the initial brane configuration which are put in by hand (or by "principles"). This picture also involves a singularity (when the branes collide, also meant to be the starting point of the standard big bang) and considerable controversy surrounds the questions of how this singularity might be resolved (see for example Kallosh et al. 2001, Khoury et al. 2001b, and Gordon and Turok 2002). 4) In terms of predictions, much depends on how the singularity is resolved. Certainly the homogeneity and flatness of the universe (predictions of inflation) are put by hand into the initial conditions of this model. Some argue that predictions for cosmic perturbations in these models have already ruled them out (Tsujikawa et al. 2002), but others argue that the predictions are consistent with what we know so far, but offer novel differences from inflationary predictions that could be observed in the future (Khoury et al. 2002).

Because of the differences on points 1,2 and 4, the Ekpyrotic universe does not represent an alternative mechanism that can replace inflation by doing what inflation does in a different way. As far as initial conditions are concerned it is, much like eternal inflation, a retreat back to the conceptual framework of a stand-alone standard big bang, where most of the specifics of the state of the universe are put into the initial conditions by hand. However, as with eternal inflation, if the vision of the original authors pans out this idea will offer a resolution of the big bang singularity. In addition, the Ekpyrotic idea suggests intriguing testable predictions for cosmic perturbations.

**The Cyclic Universe:** If the singularity of the ekpyrotic universe *can* be resolved in the manner originally proposed, very similar dynamics could also be used to construct a cyclic model of the universe (Steinhardt and Turok 2002a). Although some notion of a cyclic universe has been around for a long time (Tolman 1937), suitable dynamics to turn a contracting universe into an expanding one were always lacking. If the brane-collision picture can be shown to work, it will offer a nice way to construct a universe that bounces from contraction back into expansion. Using this innovation, Steinhardt and Turok (2002a) constructed a cyclic model of the universe which includes a period of inflation late in the cycle. Rather efficiently, this proposal uses today's cosmic acceleration[13] as the inflation period for the next cycle. With a period of inflation built into the scenario one might be tempted to view the cyclic universe as a variation on the inflation theme, and indeed, modulo clarifying what happens at the singularity, I regard this as a pretty interesting variation.

However, Turok and Steinhardt originally state that a key feature of the new cyclic scenario was to offer completely eternal cyclic evolution (see for example Steinhardt and Turok (2002b)). In this picture, like eternal and ekpyrotic scenarios discussed above, there is no pre-inflation state to contend with. One simply declares "this is the state of the universe".

---

[13] For a review of the cosmic acceleration see for example (Albrecht 2002)

My discussion in this article leads to a number of concerns about this claim of eternality. First of all, the claim of eternality is a very extreme one. If there is any non-zero probability, no matter how small, of the model fluctuating (unstably) off its cycle, that fluctuation has all of eternity to get around to happening, and thus it is 100% certain to happen at some time. Any such event will completely destroy any claims to eternality.

The arrow of time is a nice illustration of just this sort of effect. While to some approximation the arrow of time can be regarded as an absolute property of our physical world, a deeper analysis reveals the arrow of time indeed to be only an approximation. Just as it is possible for air to spontaneously rush into one corner of a room, and just as in fact such a rare event is absolutely certain to happen if you wait long enough, there are probably many different ways some "conspiracy" of microscopic degrees of freedom could conspire to divert the oscillating universe from its cycle. To make a compelling case for eternality, one would have to argue that all possible rare events had been completely accounted for. Such a case has certainly not been made so far, and it is very hard to see how such a case ever could be made.

This issue must in some sense be a weakness of the eternal and ekpyrotic scenarios as well, but in those models one controls more aspects of the state of the universe simply by declaration (namely making eternality part of the definition of the state). The new cyclic model is presented in a way that leaves more details in the hands of dynamical evolution (possible because of the attractor behavior during the regular periods of inflation). I feel this greater focus on dynamics is a strength of the cyclic model, but it also makes it easier to formulate the concern that a very rare event could prevent eternality.

To be more specific, one can study the origin of the arrow of time in the cyclic model. A crucial role is played by the assumption that heat (and entropy) is reliably produced upon brane collision but the time reverse (cooling) *never* occurs. In the current literature this feature is put in completely "by hand" and only at the "thermodynamic" level. Namely, the current treatment uses what is effectively a friction term to impose an arrow of time on the cyclic model. Just as a deeper understanding of everyday friction allows for the ridiculously small but non-zero probability that a coherent fluctuation could appear and produce a push in the opposite direction to normal friction, one would expect whatever microscopic mechanism that underlies the friction term in the cyclic universe would be able to do the same. Because eternality is such an extreme claim, one such fluctuation could be enough to destroy eternality. In any case, it would certainly be interesting to learn what microscopic picture the advocates of the eternally cycling model have in mind.

I should reiterate, my criticism of claims of eternality does not detract from the appeal the cyclic model could have as a new mechanism within the normal conceptual framework of inflation, with the arrow of time originating as a rare fluctuation[14].

**Varying Speed of Light (VSL):** Another approach to initial conditions is based on the idea that the speed of light could have been faster in the past (Albrecht and Magueijo 1999, Moffat, 1993). These models are still in their early stages and a clear picture of the fundamental origin of the varying speed of light, as well as the origin of perturbations, has yet to emerge. Still the VSL concept attempts to duplicate the approach of inflation on all four points discussed in this section. In particular one would expect any fundamental theory that allowed the speed of light to vary would make it just as likely to be slower or faster in the past. In that sense the speed of light would have to play a similar role to the inflaton, linking the arrow of time to a rare fluctuation in *c(t)*.

---

[14] I recently learned that authors of the new cyclic model no longer see eternality as an important goal of the model and agree that the cyclic universe is unlikely to be truly eternal (Steinhardt and Turok 2002c).

**Holographic Cosmology:** Another intriguing proposal uses the idea of holographic bounds on the entropy of gravitating systems to describe a maximal entropy "black hole gas" state from which our big bang universe emerged (Banks and Fischler 2001). In this work, Banks and Fischler take the view that the causal structure around an observer is absolutely fundamental, and build a physical picture on top of that. The arrow of time appears in their picture as a fundamental feature (not an emergent or approximate one) linked to the causal structure. Thanks to space of states of matter that actually grows with time, the dynamics in this picture are not even reversible at the microscopic level. In this way the holographic cosmology picture is completely different from the standard inflationary picture discussed here.

Still, there are some interesting parallels. In particular, the global properties of the black hole gas state are very different from the universe we observe, so Banks and Fischler propose a dominant channel (quite different from inflation) whereby rare regions in the black hole gas evolve into something like the standard big bang. Much still needs to be developed in this picture, particularly the origin of cosmic structure on large scale, but already it offers the most dramatic and stimulating departure yet from standard ideas about cosmic initial conditions.

**Wavefunction of the universe:** Many have been tempted to think that some argument or principle could define the "wavefunction of the universe" (Hartle and Hawking, 1983; Linde 1984; Vilenkin 1984, Hawking and Turok 1998, Linde 1998, Hawking and Hertog (2002))[15]. So far such attempts have yielded different wavefunctions in the hands of different authors (Vilenkin 1998). On the face of it this approach is fundamentally different from the inflation-based picture discussed in this article. Simply declaring the wavefunction of the universe is not about dynamical mechanisms that give a preferred channel from chaos into the standard big bang. It is about principles simply choosing the state of the universe by one method or another, much like we see in the eternal and ekpyrotic cases.

Interestingly, many discussions include both the wavefunction of the universe and inflation, and use the wavefunction of the universe to determine the most probable way that inflation gets started. In that work, the wavefunction of the universe is basically an approach to describing the pre-inflation state. Most proposed wavefunctions of the universe are not all that sharply peaked, and their breadth might be interpreted as an expression of the "chaos" often assumed for the pre-inflation state, perhaps tempered slightly by some general principles. If things develop in this way, the wavefunction of the universe idea might turn out to be more closely connected to standard ideas about inflation than superficially appears to be the case.

Another interesting idea advocated in some wavefunction of the universe discussions is that classical spacetime must be automatically correlated with an arrow of time. In these discussions the wavefunction of the universe is used to provide constraints on classical spacetime as it emerges from a quantum regime, and it is argued that the classical

---

[15] Of course the starting point is usually the Wheeler-DeWitt equation (Dewitt, 1967; Wheeler, 1968)

spacetimes that emerge naturally come with low and high entropy "ends" and thus an arrow of time (Further discussion of this point of view with additional references can be found in Zeh (1992), Gell-Mann and Hartle (1994) and Hawking (1994).) So far these arguments are made in the context of "mini-superspaces" based on a FRW background, which of course pre-supposes the homogeneity that accounts for the actual arrow of time in the universe. If these results persist in a more complete theory (with a more complete superspace) this line of reasoning could give key insights into origin of the arrow of time. In such a picture one could predict the arrow of time by simply using the classicality of spacetime as an observational input.

Of course, it is not at all clear that stating the wavefunction of the universe from first principles will ever take hold as a theory of initial conditions. The physical world clearly has a huge phase space which it tends to explore in a thorough way, and I am far from convinced that principles concocted by humans could really convince the universe to avoid large parts of that phase space. My skepticism is only enhanced by the fact that we do not have one theory of the wavefunction of the universe, but many, and the community as a whole has not found compelling reasons to choose one over the others.

**Chaotic Mixing:** Cornish et al. (1996) argue that "chaotic mixing" that can occur in topologically complex spaces could dynamically "explain" the initial conditions for a special type of inflation. This line of reasoning seems to be in conflict with the idea that there has to be something non-generic or rare about the initial conditions for inflation in order to have an arrow of time. As far as I can tell, this idea has not been well enough developed to determine where the arrow of time fits in. There seems to by an asymmetric treatment of the "instability" associated with inflation it also appears that the usual gravitational instability is not accounted for at all in their discussion.

Another somewhat related idea proposes that the homogeneity of the early universe arises from a kind of statistical averaging over higher dimensions (Starkman et al. 2001). In that model, the arrow of time is put in by hand via the assumptions about the initial state. They assume FRW topology, as well as a statistical ensemble of states with an average curvature of zero (far from gravitational collapse). Both these assumptions produce an initial state which is effectively "low entropy", and thus generates an arrow of time. The authors argue that this is an explanation of the homogeneity of the universe, but it is hard to imagine this dynamical mechanism competing with inflation. The statistical ensemble of states they require, with no mean curvature over a huge volume, seems to be much lower entropy (and thus much more rare) than the small inflaton fluctuation required for inflation. Also, unlike inflation, Starkman et al.'s mechanism does not generate any large volume factors which could leverage their mechanism.

**Brane Gas Cosmology:** This idea proposes that the homogeneity of our universe emerges from some kind of equilibration process of branes in higher dimensions (Watson and Brandenberger 2002). As in the cases of Chaotic Mixing and Holographic cosmology, it is not clear that care has been given to arrow of time issues in this model. Something has to fluctuate or "be declared" out of equilibrium in order to have the arrow of time we observe. What degree of freedom takes on that role in the Brane Gas model?

**Cardassian Expansion:** The "cardassian expansion" model (Freese and Lewis 2002, Freese 2002) comes from modified Friedman equations which have homogeneity built in, and as such does not address the origin of homogeneity in the universe. Unless this model develops into one that does address the homogeneity of the universe (the origin of the arrow of time) it is not possible to analyze the relationship of the cardassian model to the arrow of time and other topics discussed here.

## 8 Further Discussion

### 8.1 Emergent Time and Quantum Gravity

Microscopic time has long been regarded as a problematic notion for a full theory of quantum gravity. One attractive resolution of the problem of time in quantum gravity is to view microscopic time as an emergent quantity that does not have to be well defined for all states of the universe. Operationally speaking, microscopic time is just a statement about correlations between physical systems designated as clocks, and other physical systems of interest. Perhaps we should understand quantum theory most fundamentally as a theory of correlations. These correlations can only be organized according to a microscopic time parameter under conditions where physical subsystems exist that actually behave like good clocks. Depending on what the complete space of states looks like for a full theory of quantum gravity, states with "good clock" subsystems and thus well-defined microscopic time might only be a small subset of all possible states. (For reviews and further references on the problem of time in quantum gravity including the notion of emergent time, see Kuchar (1992), Isham (1993) and chapter 6 of Zeh (1992). See also Albrecht (1995).)

If microscopic time *is* emergent, how could that affect the discussion in this article, which for the most part is basically classical? Perhaps not at all: To discuss time, of course one has to restrict oneself to physical states where microscopic time is well defined. But having done so in the context of a full theory of quantum gravity, one may well be faced with the exact situation discussed classically in this article, namely one in which almost all possible states do not have a thermodynamic arrow of time, and one has to make an additional selection to identify those special states which do.

Another possibility is that microscopic time emerges pre-packaged with a thermodynamic arrow of time, so that "good clock subsystems" naturally come correlated with matter states that are very low entropy at one end of the timeline and high entropy in the other direction. This line of reasoning appears in many discussions of the wavefunction of the universe in the context of quantum cosmology: (See the "Wavefunction of the Universe" discussion in section 7).

Of course another alternative is that once we have a full understanding of quantum gravity we will learn that all states have a well-defined microscopic time, which comes automatically correlated with a thermodynamic arrow of time (see for example the "Holographic Cosmology" discussion in section 7).

### 8.2 Causal Patch Physics

One popular explanation of today's observed cosmic acceleration posits a non-zero value of the fundamental cosmological constant which is today just starting to dominate over the energy density in ordinary matter. The acceleration might be thought of as a second period of inflation, but if it is driven by a fundamental cosmological constant then unlike inflation the acceleration today will not come to an end.

A non-zero cosmological constant could revolutionize fundamental physics (Banks, 2000, 2002; Fischler, 2000; Witten 2000). In particular, Banks and Fischler argue that a cosmological constant would place an absolute upper bound on the entropy of the universe, which in turn would imply a finite dimensional Hilbert space for any fundamental theory. This is related to the fact that one only assigns physical meaning to events to which you are causally connected, that is, your "causal patch".

Dyson et al. (2002) have explored the implications of this idea for inflation. The good news is that with a fundamental cosmological constant, some things become simpler. For example, the highest entropy state in such a universe is pure de Sitter space, and so it must be de Sitter space that describes the "pure chaos" that preceded inflation. This situation appears to be theoretically much more tractable than trying to conceive of the perfectly chaotic state in the absence of a cosmological constant.

But challenges arise if one embraces the finite Hilbert space idea. The first of course, is that no one knows a compelling fundamental theory that fits this constraint. But Dyson et al. (2002) argue that even without those details, the causal patch constraint deprives inflation of its huge volume factors. In this picture, the entire volume of the universe is not much larger that what we see, and without the usual exponentially large volumes, Dyson et al. argue that inflation is *not* the dominant channel into the standard big bang.

So one must abandon either inflation, the non-zero cosmological constant, or the causal patch constraint (at least in the heuristic form use by Dyson et al.).

Turok has also argued the case for excluding the large volume factors using an argument based on causality, without specific reference to a cosmological constant (Turok 2002a), and similar concerns are raised by Hawking and Hertog (2002).

*8.3    Measures and other issues*

I should acknowledge that much of the discussion of what inflation has to offer (for example that the large volume factors make inflation the dominant channel from chaos into the big bang) rests on very heuristic arguments. The program of putting these sorts of arguments on firmer foundations is in its infancy. Also, there are many poorly developed technical matters related to the origin of perturbations in inflation (see for example Brandenberger and Martin 2002 and Kaloper et al. 2002). Progress on these issues certainly has the potential to overturn many of the beliefs about inflation expressed in this article.

# 9    Some "Wheeler Class" questions

John Wheeler has never shied away from the really tough "big questions". In his honor, I take this section to touch on some deeper questions raised by my discussion above. At this point, I cannot offer answers to any of these questions.

*9.1    The arrow of time, classicality and microscopic time*

In section 3.3 I mentioned the key role played by the arrow of time in quantum measurement. If one models the universe in a way where the arrow of time is only a transient phenomenon, what then do quantum probabilities mean in the absence of the arrow of time? We seem pretty comfortable working with such probabilities, but perhaps we should be more careful here. (For a recent discussion of some of these issues, see Banks *et al* (2002).)

Also, we are used to thinking of the time that appears in the microscopic equations (and which is differentiated from space thanks to its different role from space in Lorentz transformations) as being quite different from the arrow of time under discussion here. For example, it is the microscopic arrow of time that allows us to construct a time sequence for a box of gas in equilibrium (such as depicted in Figure 2) even when a thermodynamic arrow of time does not exist.

If one thinks carefully about how one operationally defines this microscopic time, the thermodynamic arrow of time is always required indirectly. One might say something like: "measure a system at time 1, and the microscopic evolution equations will tell you what the system will look like at time 2". But to actually check that you have to make a good record of the state at time 1, a record that will still be intact at time 2. As discussed in section 3.3, the thermodynamic arrow of time is essential to making stable records.

So perhaps one cannot really have microscopic time, or even quantum probabilities, without a thermodynamic arrow of time. This idea might connect with other speculation that microscopic time, like the thermodynamic time, could be an emergent feature of the physical world (see section 8.1 and (Gross 2002)). On the other hand, the ideas from "Holographic Cosmology" (discussed in section 7) offer a very different angle that also connects microscopic time with the arrow of time in a fundamental way.

### 9.2    *The arrow of time in the approach to de Sitter space*

To us, the familiar arrow of time is driven by gravitational collapse as it destroys a homogeneous state. In many models of inflation there are huge regions of the universe that are undergoing extremely long epochs of exponential expansion. As time progresses and various imperfections get diluted by the expansion, these regions will asymptotically approach the high entropy de Sitter state, and by doing so will exhibit increasing entropy. That is another manifestation of an arrow of time, which is quite different from the one we are used to which is based on gravitational collapse, not dilution. Is it possible for other types of creatures to exist that harness the "dilution" arrow of time as effectively as we harness ours? If the answer is yes, perhaps these creatures will start to evolve as today's cosmic acceleration takes over.

## 10   Conclusions

Perhaps the key point of this article is that having an arrow of time in the universe places demands on the initial conditions that apparently conflict with the goals of inflationary cosmology. Inflation wants to use dynamics to argue that the initial conditions of the big bang are generic, but the arrow of time requires that the initial conditions not be generic in precisely the same sense, namely that the conditions are far from the asymptotic behavior produced by the dynamics.

This conflict is resolved by recognizing that the goal of inflation can never be to make the initial conditions of our observed universe be *completely* generic. To do so would remove the arrow of time. However, inflation teaches us that it *is* possible to make the initial conditions of our universe "more generic": The inflationary dynamics shows that the special initial conditions required for an arrow of time need only appear as special initial conditions for the inflaton field in a small region, not the entire state of matter in the universe. Inflationary dynamics then leverages these special inflaton initial conditions into exponentially large numbers of exponentially large regions that exhibit the properties of the familiar big bang.

So one important conclusion is that inflation, or any other attempt to dynamically explain the initial conditions of the observed universe, will necessarily require some special initial conditions itself, in order to have an arrow of time. These special initial conditions are the vestiges of Boltzmann's original "rare fluctuations" which can never be completely excised from this sort of dynamical approach.

This conclusion is particularly directed at those who hold up the special initial conditions of inflation as a serious flaw of the idea. However, I know of no fundamental law that prevents one from hoping that some improved dynamical process could produce the universe we observe using initial fluctuations that are even less rare than those which initiate inflation, so perhaps it is just as well that the critics keep the pressure on.

But the discussion in this article also relates to another debate about initial conditions. There are those who find the dynamical approached inherently flawed. Instead, they wish to uncover broad principles or fundamental laws that will uniquely specify the state of the universe (see for example Hollands and Wald, 2002). As I discussed in section 7, several current ideas (such as the ekpyrotic model and eternal inflation) fall under this category. The field seems to be divided among people who strongly favor a dynamical approach, and those who strongly favor defining a unique state of the universe based on principles. I am definitely in the dynamical camp.

My most concrete criticism of the "unique state" approach is that all the dynamics we know, especially when quantum effects are included, tends to spread states out in phase space in a very broad manner. I challenge proponents of the unique state approach to articulate their ideas in a fully quantum treatment. I suspect that any such attempt will yield a probability distribution for the state of the universe that is broad in many directions, effectively describing something not that different from the pre-inflation chaos discussed above.

Under those conditions, there is no other way of finding our place in the universe besides identifying the dynamical mechanisms that are most likely to produce the big bang universe we observe. As I have argued here, the fact that we experience an arrow of time requires that these dynamical mechanisms also need some kind of rare fluctuation to function, a feature that is deeply connected with Boltzmann's original insights into the arrow of time more than a century ago.

**Acknowledgements**

I would like to thank A. Aguirre, T. Banks, S. Carroll, W. Fischler, N. Kaloper, M. Kaplinghat, A. Linde, A. Olinto, L. Susskind, N. Turok, and W. Zurek for very helpful conversations. I also thank the Aspen Center for Physics, where much of this article was written, and B. Gold for comments and corrections to the manuscript.